\documentstyle[twoside,fleqn,npb,epsfig]{article}
\newcommand{\AmS}{{\protect\the\textfont2
  A\kern-.1667em\lower.5ex\hbox{M}\kern-.125emS}}

\title{The NNLO QCD analysis of the CCFR data for $xF_3$ :
is there still the room for the twist-4 terms ?}

\author{A.L. Kataev\address{Institute for Nuclear Research of the
        Academy of Sciences of Russia,\\
        117312 Moscow, Russia}
        \thanks{Supported in part by the Russian Foundation
        of Basic Research, Grant N 99-01-00091},
	G. Parente\address{Department of Particle Physics, University
	of Santiago de Compostela,\\
       15706 Santiago de Compostela, Spain}\thanks{Supported by
       CICYT (AEN96-1773) and Xunta de Galicia (Xuga-20602B98)}
        and
        A.V. Sidorov\address{Bogoliubov Laboratory of Theoretical Physics,
        Joint Institute for Nuclear Research, \\
        141980 Dubna, Russia}\thanks{Supported in part by the Russian Foundation
        of Basic Research, Grant N 99-01-0091 and by the Heisenberg-Landau
        program.}}

\begin{document}
\begin{abstract}
The results of the more detailed NNLO
QCD analysis of the CCFR data for $xF_3$ SF are presented.
The factorization scale uncertainties are analyzed. The  NNLO
results for $\alpha_s(M_Z)$ and twist-4 contributions are obtained.
Despite the fact that the amplitude of the
$x$-shape of the twist-4 factor is consequently decreasing at
the NLO and NNLO, our new QCD analysis seems to reveal
the remaining twist-4 structure at the NNLO level. The definite
N$^3$LO uncertainties  are fixed using
the [0/2] Pad\'e resummation technique.
\end{abstract}
\maketitle
It is known that the CCFR collaboration
provided not long ago rather precise experimental data for $xF_3$
SF of $\nu N$ DIS and extracted the value
of $\alpha_s(M_Z)$ using the NLO DGLAP analysis\cite{CCFR}.
In its process the twist-4 contributions were taken into account
using the infrared renormalon (IRR) model of Ref.\cite{DW} with its
parameter, fixed as 1/2 of the originally proposed one. 

In the series of the subsequent papers\cite{KKPS1}-\cite{KPS}
we concentrated  on the attempts to fit the CCFR data at the
NNLO level with the help of the Jacobi polynomial-Mellin moments
version of the DGLAP method \cite{PS}-
\cite{Kriv}, based on the following equation:
\begin{eqnarray}
&&xF_3^{N_{max}}=x^{\alpha}(1-x)^{\beta}\sum_{n=0}^{N_{max}}
\Theta_n^{\alpha,\beta}(x) \times \\ \nonumber
&& \sum_{j=0}^{n}c_j^{(n)}(\alpha,\beta)
M_{j+2}(Q^2)+\frac{h(x)}{Q^2}
\end{eqnarray}
where $h(x)$ is the twist-4 contribution.

We used the results
of calculations of  the NNLO corrections to the coefficient functions\cite{VZ}
and the  analytical expressions for the NNLO corrections
to the anomalous dimensions of the nonsinglet moments with
$n=2,4,6,8,10$ \cite{LRV}, supplemented with the given in Ref.\cite{KKPS1}
$n=3,5,7,9$ similar numbers, obtained using the smooth interpolation
procedure of Ref.\cite{PKK}.

Using the
fits with free Jacobi polynomial parameters $\alpha,\beta$,
we found that their values 
$\alpha\approx 0.7$, $\beta\approx 3$
corresponds to  the minimum in the plane $(\alpha,\beta)$. The
form of $h(x)$ was fixed (1) through the IRR model
of Ref.\cite{DW} with its coefficient $A_2^{'}$ considered as the
free parameter and (2) as the function, modeled by  free parameters 
$h_i=h(x_i)$, where $x_i$ are
the points of the experimental data bining. The QCD evolution of
the moments has the following form
\begin{equation}
\frac{M_n(Q^2)}{M_n(Q_0^2)}=
\bigg(\frac{A_s(Q^2)}{A_s(Q_0^2)}\bigg)^{\frac{\gamma_{NS}^{(0)}}{2\beta_0}}
\frac{AD(n,Q^2)C^{(n)}(Q^2)}{AD(n,Q_0^2)C^{(n)}(Q_0^2)}
\end{equation}
where $A_s(Q^2)=\alpha_s(Q^2)/(4\pi)$ is the $\overline{MS}$-scheme
expansion parameter, $AD(n,Q^2)=1+p(n)A_s(Q^2)$ $+q(n)A_s(Q^2)^2+...$
comes from the   
\begin{table*}[hbt]
\setlength{\tabcolsep}{1.5pc}
\newlength{\digitwidth} \settowidth{\digitwidth}{\rm 0}
\catcode`?=\active \def?{\kern\digitwidth}
\caption{
The $Q_0^2$ dependence  of  
$\Lambda_{\overline{MS}}^{(4)}$ [MeV].
$LO^{*}$ means that in the
$LO$-fits $NLO$  $\alpha_s$
is used; $NLO^{*}$ ($NNLO^{*}$) indicates that
in the $NLO$ ($NNLO$) fits $NNLO$ ($N^3LO$) 
$\alpha_s$ is used. $[0/2]$ marks the results
of  the $N^3LO$ expanded $[0/2]$ Pad\'e fits
with $\alpha_s$ defined at the $N^3LO$.
}
\label{tab:effluents}
\begin{tabular*}{\textwidth}{@{}l@{\extracolsep{\fill}}rrrrrr}
\hline
$Q_0^2(GeV^2)$& 5  &     8 &       10    &  20  &     50   &    100    \\
\hline
   $LO$    &   266  $\pm$35 & 266$\pm$35&  265$\pm$34 & 264$\pm$35&  264$\pm$36 & 263$\pm$36  \\
   $LO^{*}$   &   382  $\pm$38 & 380$\pm$41&  380$\pm$40 & 379$\pm$46&  378$\pm$43 & 377$\pm$42  \\
   $NLO$   &   341  $\pm$30 & 340$\pm$40&  340$\pm$35 & 339$\pm$36&  337$\pm$34 & 337$\pm$37  \\
   $NLO^{*}$  &   322  $\pm$29 & 321$\pm$33&  321$\pm$33 & 320$\pm$34&  319$\pm$36 & 318$\pm$36  \\
   $NNLO$  &   293  $\pm$30 & 312$\pm$33&  318$\pm$33 & 326$\pm$35&  326$\pm$36 & 325$\pm$36  \\
\hline
   $NNLO^{*}$ &   284  $\pm$28 & 303$\pm$31&  308$\pm$32 & 316$\pm$33&  316$\pm$33 & 315$\pm$34  \\
  $ [0/2]$ &   293  $\pm$29 & 323$\pm$32&  330$\pm$35 & 335$\pm$37&  326$\pm$36 & 319$\pm$35  \\
\hline
\end{tabular*}
\end{table*}
\\[-10mm]
expansion of anomalous dimesnion,
$C^{(n)}(Q^2)=1+C^{(1)}(n)A_s(Q^2)+C^{(2)}(n)A_s(Q^2)^2+...$ is
the coefficient function of $M_n(Q^2)$ and $M_n(Q_0^2)$
is parametrized
at the factorization scale $Q_0^2$.
In the case of $f=4$ numbers of flavours the numerical values  of  $p(n)$, $q(n)$, $C^{(1)}(n)$
and $C^{(2)}(n)$ 
are given in Ref.\cite{KPS}. We will use the expansion
 of $\alpha_s$ through the  powers of
$1/ln(Q^2/\Lambda_{\overline{MS}}^2)$ in the LO, NLO, NNLO and
N$^3$LO,  which contains the 
4-loop term of the QCD $\beta$-function \cite{RVL}. 

Here we complete  previous  analysis of the CCFR data of
Refs.\cite{KKPS2,KPS}, performed in the case of 
$Q_0^2=5~GeV^2$, by varying
$Q_0$ in the wide  region. 
The  fits were done for the CCFR data, cutten at $Q^2>5~GeV^2$,
without  twist-4 effects, 
but with target mass corrections included. The
results for
$\Lambda_{\overline{MS}}^{(4)}$  
are given in Table 1
for different  $Q_0^2$.
The  LO and NLO results are
stable to the choice of $Q_0^2$.
The results of the
$LO^{*}$ fits are higher than the LO ones, and from this
level taking into account of  other perturbative QCD effects
are decreasing the  values of
$\Lambda_{\overline{MS}}^{(4)}$.
The NNLO results are  sensitive
to the variation  of $Q_0$. 
The values of $\Lambda_{\overline{MS}}^{(4)}$ become
stable for $Q^2\geq 20~GeV^2$ only. The same effect is manifesting
itself for the results of the $N^3LO$ fits.
This effects can be related to the
peculiar behavior of the NNLO perturbative QCD expansion
of $M_2$. Using the 
numerical values of  $p(2)$, $q(2)$, $C^{(1)}(2)$
and $C^{(2)}(2)$, given in Ref.\cite{KPS}, we obtain 
\begin{eqnarray}
&& AD(2,Q^2)C^{(2)}(Q^2)=1-0.132A_s(Q^2) \\ \nonumber
&& -46.155A_s(Q^2)^2+...
\end{eqnarray}
Thus,  it is safer 
to start the evolution from 
$Q_0^2=20~GeV^2$, where the numerical value
of the $A_s^2$ contribution in Eq.(3) is smaller. 

The results of our new fits of the CCFR data 
with twist-4 contributions,
fixed through the IRR model of Ref.\cite{DW}, are presented
in Table 2. 
\\ [-10mm]
\begin{table}[hbt]
\catcode`?=\active \def?{\kern\digitwidth}
\caption{
 The results for
$\Lambda_{\overline{MS}}^{(4)}$ and
the IRR coefficient $A_2^{'}$ [ $GeV^2$] from 
the CCFR'97 data, cutten at $Q^2>5~GeV^2$ in different orders 
for  $Q_0^2$=$20~GeV^2$.}
\label{tab:effluents1}
\begin{tabular}{lrcr}
\hline
                $  $           &
                $\Lambda_{\overline{MS}}^{(4)}$ (MeV)     &
                $ A_2^\prime$(HT)   &
                $ \chi^2$/points   \\
\hline
LO   & 433$\pm$52      & -0.33$\pm$0.06 & 82.8/86      \\
NLO   & 369$\pm$45      & -0.12$\pm$0.06 & 81.8/86      \\
NNLO   & 326$\pm$35      & -0.01$\pm$0.05 & 76.9/86      \\
N$^3$LO   & 340$\pm$37      & -0.04$\pm$0.05 & 77.2/86      \\
\hline
\end{tabular}
\end{table}
\\[-10mm]
When the twist-4 parameters 
are not taken into account, the effects of 
the NNLO corrections are smaller than in the 
case of our previous analysis of Refs.\cite{KKPS2,KPS}
(see Table 1).
However, they
are  still sizable in the case when twist-4
contributions are fixed through the IRR model. 
They 
have the tendency to make the value
of  $A_2^{'}$ 
comparable with zero.
As the result, the NNLO value of
$\Lambda_{\overline{MS}}^{(4)}$ is the same in the cases
of both neglecting and retaining twist-4 terms.

At Fig.1 we present the extraction of
the $x$-shape of the twist-4 terms from the LO, NLO,
NNLO and expanded Pad\'e fits with $Q_0^2=20~GeV^2$
for the unfixed twist-4 contribution.
For    
$\Lambda_{\overline{MS}}^{(4)}$ we got:
331$\pm$162 MeV (LO level),
440$\pm$ 183 MeV (NLO level),
372$\pm$ 133 MeV (NNLO level)
and 371 $\pm$ 127 MeV ( expanded [0/2] Pad\'e) 
\\ [-15mm]
\begin{figure}[htb]
\vspace{-20pt}
\medskip
\parbox[b]{7cm}{\psfig{width=6cm,file=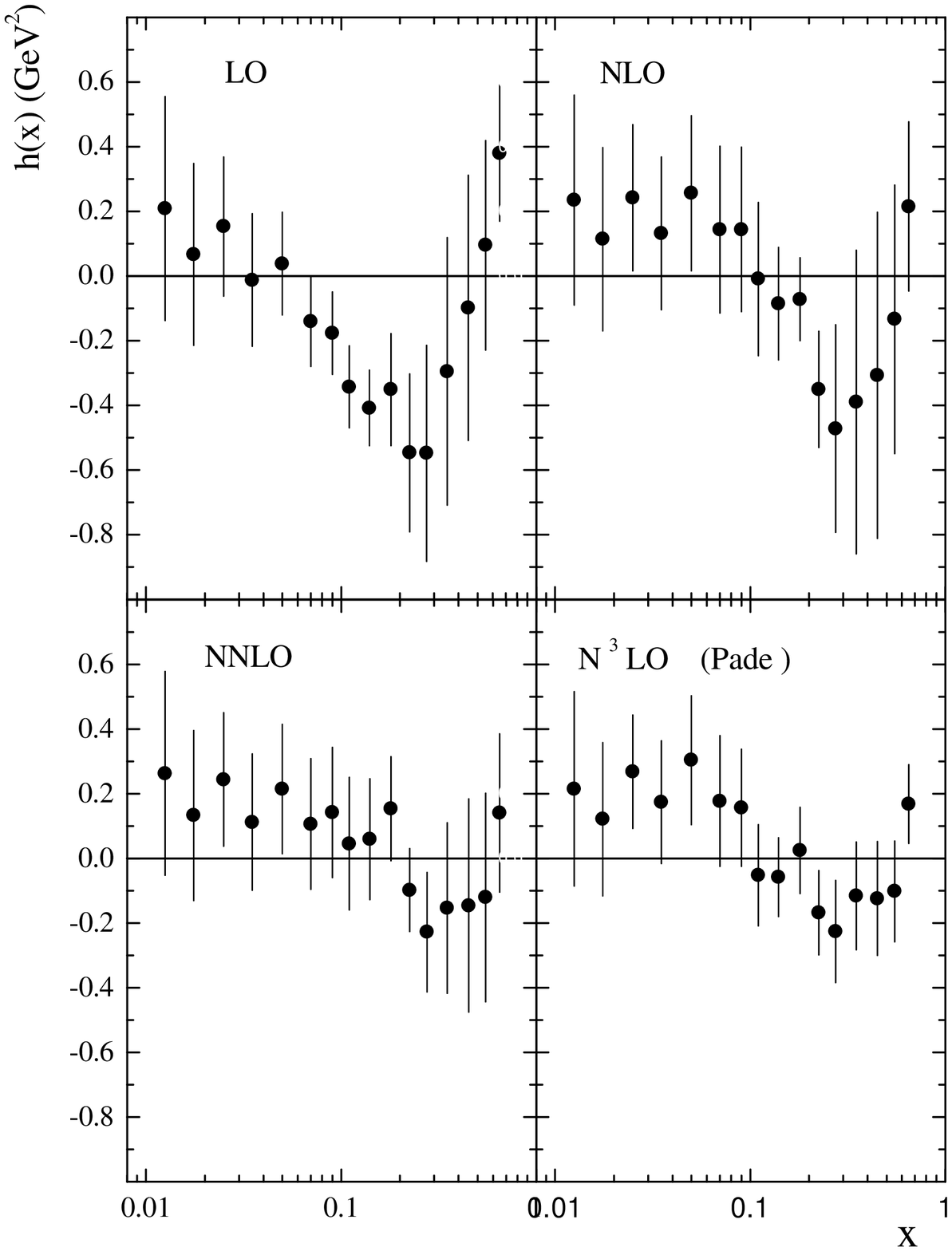}} \\ [-25mm]
\caption{The $x$-shape of $h(x)$ extracted from the
fits of CCFR'97 for $Q_0^2=20~GeV^2$.}
\label{fig:largenenough}
\end{figure}
\\[-10mm]
One can see
that taking into account of the 
higher order perturbative  corrections is
decreasing the amplitude of the variation of $h(x)$. This observation is
in agreement with the results 
of Refs.\cite{KKPS2,KPS}, obtained for the case of $Q_0^2=5~GeV^2$.
However, the change of factorization scale 
allows  
to detect the remaining twist-4 structure even at the NNLO.
It is  relatively stable to 
the application of the 
$[0/2]$ Pad\'e approximations method.

When the twist-4 terms are fixed through the IRR model 
we obtain
\begin{eqnarray}
\alpha_s(M_Z)|_{NLO}&=&0.120 \pm 0.002 (stat)\\ \nonumber
&&\pm 0.005 (syst)\pm 0.004 (th.)
\\ \nonumber 
\alpha_s(M_Z)|_{NNLO}&=&0.118 \pm 0.002(stat)\\ \nonumber 
&&\pm 0.005 (syst) \pm 0.003 (th.)
\end{eqnarray}
and
\begin{eqnarray}
\alpha_s(M_Z)|_{NLO}&=& 0.123^{+0.008}_{-0.010} (stat) \\ \nonumber 
&& \pm 0.005 (syst) \pm 0.004 (th.) \\ \nonumber
\alpha_s(M_Z)|_{NNLO}&=& 0.121^{+0.007}_{-0.009} (stat) \\ \nonumber 
&& \pm 0.005 (syst) \pm 0.003 (th)
\end{eqnarray}
when the twist-4 terms  
parameters $h(x_i)$ are free.
The systematic errors are taken from the CCFR experimental analysis and 
the theoretical (th.) ambiguties  are 
dominated by the uncertainty in the choice of the matching point
in the NLO, NNLO and N$^3$LO variants of the 
$\overline{MS}$-matching condition \cite{ChKS}, derived 
following the lines of Ref.\cite{BW}.
It was estimated by varying  
$b$-quark threshold from $M_b=m_b$ to $M_b=6.5m_b$ \cite{BV} and 
is of over $\pm 0.002$.


{\bf Acknowledgements.} We are grateful to G. Altarelli,
J. Bl\"umlein and W. van Neerven for the comments on 
the previous results \cite{KKPS2,KPS} of our
research. One of us (ALK) wishes thank the OC of DIS99 Workshop 
for the invitation and financial support.


\begin{thebibliography}{99}
\bibitem{CCFR} CCFR-NuTeV Collab., W.G. Seligman et al.,
{\it Phys.Rev.Lett.} {\bf 79} (1997) 1213.
\bibitem{DW} M. Dasgupta and B.R. Webber, {\it Phys.Lett.}
{\bf B382} (1996) 273; 
M. Maul et al.,
{\it Phys. Lett.} {\bf B401} (1997) 100.
\bibitem{KKPS1} A.L. Kataev, A.V. Kotikov, G. Parente and
A.V. Sidorov, {\it Phys. Lett.} {\bf B388} (1996) 179.
\bibitem{KKPS2} A.L. Kataev, A.V. Kotikov, G. Parente and
A.V. Sidorov, {\it Phys. Lett.} {\bf B417} (1998) 384.
\bibitem{KPS} A.L. Kataev, G. Parente and A.V. Sidorov,
hep-ph/9809500.
\bibitem{PS} G. Parisi and N. Sourlas, {\it Nucl. Phys.}
{\bf B151} (1979) 421. 
\bibitem{ChR} J. Ch\'yla and J. Ramez, {\it Z. Phys.}
{\bf C31} (1986) 151.
\bibitem{Kriv} V.G. Krivokhizhin et al., {\it Z. Phys.}
{\bf C36} (1987) 51;
{\it Z. Phys.} {\bf C48} (1990) 347.
\bibitem{VZ} E.B. Zijlstra and W.L. van Neerven, {\it Nucl. Phys.}
{\bf B417} (1994) 61.
\bibitem{LRV} S.A. Larin, T. van Ritbergen and J.A.M. Vermaseren,
{\it Nucl. Phys.} {\bf B427} (1994) 41; S.A. Larin 
et al., 
{\it Nucl. Phys.}
{\bf B492} (1997) 338.
\bibitem{PKK} G. Parente, A.V. Kotikov and V.G. Krivokhizhin,
{\it Phys. Lett.} {\bf B333} (1994) 190.
\bibitem{RVL} T. van Ritbergen, J.A.M. Vermaseren and S.A. Larin,
{\it Phys. Lett.} {\bf B400} (1997) 379.
\bibitem{ChKS} 
K.G. Chetyrkin, B.A. Kniehl and M. Steinhauser, 
{\it Phys. Rev. Lett.} {\bf 79} (1997) 2184.
\bibitem{BW} W. Bernreuther and W. Wetzel, {\it Nucl. Phys.}
{\bf B197} (1982) 228;  {\bf B513}
(1998) 758 (Err.)
\bibitem{BV}
J. Bl\"umlein and W.L. van Neerven, preprint DESY 98-176; hep-ph/9811351.
\end{thebibliography}
\end{document}